\documentclass{article}
\usepackage{amsmath}

\usepackage{amssymb}

\begin{document}

\title{A Polynomial Time Graph Isomorphism Algorithm For Graphs That Are Not  Locally Triangle-Free }
\author{ Fahad Bin Mortuza
\footnote{ \textbf{Email:}secondprime@yahoo.com}
\\Department of Computer Science And Engineering,
\\ East West University.}
\date{May 30, 2016.}
\maketitle

\begin{abstract}
In this paper, we show the existence of  a polynomial time graph isomorphism algorithm  for all graphs excluding graphs that are  \textit{locally triangle-free}. This particular class of graphs  allows to  divide the graph into \textit{neighbourhood} sub-graph where each of induced  sub-graph (neighbourhood)  has  at least $2$ vertices.  We construct all possible permutations for each induced  sub-graph using a search tree. We construct automorphisms of subgraphs based on these permutations. Finally, we decide isomorphism through  automorphisms .


\end{abstract}

\section{Introduction}
\begin{flushleft}
•
\end{flushleft}
Given two graphs $G$ and $H$, the Graph Isomorphism problem (GI) asks whether there exists a bijection from the vertices of $G$ to the vertices of $H$ that preserves adjacency. The graph isomorphism problem has a long history in the fields of mathematics, chemistry, and computing science. The problem is known to be in NP, but is not known to be in \textbf{P} or \textbf{NP-complete}. The best current theoretical algorithm is due to Babai and Luks (1983)\textbf{[2]}. The algorithm relies on the classification of finite simple groups. In 2015, Laszlo Babai  claimed  that the Graph Isomorphism problem can be solved in quasipolynomial time.
\subsection{ Notations and Definitions}
Let $G$ and $H$ be two graphs. Each graph has $n$ vertices. The \textit{cardinality} of a set   $B$ is the number of elements in it,  denoted by $|B|$. For example, $|G|=|H|=n$.

 The automorphism group of graph $G_1$ will be denoted by $Aut(G_1)$
 The \textit{neighbourhood} of a vertex $v$ in a graph $G$ is the induced subgraph of $ G$ consisting of all vertices adjacent to $v$. 

The neighbourhood is  denoted $ N_G(v)$ or (when the graph is unambiguous) $N(v)$.  If the neighbourhood does not  include  $v$  itself then it is   open neighbourhood of $v$; it is also possible to define a neighbourhood in which v itself is included, called the closed neighbourhood and denoted by $N_G[v]$. 

If all vertices in $G$ have neighbourhoods that are isomorphic to the same graph $G_1$, then $G$ is said to be \textit{locally} $G_1$, and if all vertices in $G$ have neighbourhoods that belong to some graph family $\mathcal{F},  G$ is said to be locally $\mathcal{F} $ (Hell 1978, Sedlacek 1983).

  A permutation of a vertex set $G$ is a  bijection from $G$ to itself. For example, if  $\pi=\left(\begin{matrix}  1 & 2 & 3\dots  & n  \\5& 2  & 7  \dots& 11  \end{matrix} \right)$, then the first vertex of $G$ moves to  fifth position in the $G^\pi$ . $Sym(G$) or $S_n$ denotes the set of all permutations of $G$.

 If $G$ is Isomorphic to  $H$,then $\exists P \in S_n$(Symmetric Group of $n $ vertices) such that  $H^P=G$(notation by Wielandt). We write  $G\simeq H$ when  $G$ is Isomorphic to  $H$.

A \textit{tuple} is a finite ordered list of vertices.

A \textit{search tree} is an undirected graph in which any two vertices are connected by exactly one path.A \textit{rooted search tree} is a tree in which one vertex has been designated the root.The tree elements are called \textit{nodes}.  Each of the nodes that is one graph-edge further away from a given node is called a \textit{child}, i.e  the vertices adjacent to the root vertex are called its children.A rooted tree naturally imparts a notion of \textit{Levels} (distance from the root), thus for every node a notion of children may be defined as the nodes connected to it a level below.  Nodes without children are called \textit{leaf nodes, end-Nodes, leaves}.

A \textit{walk} on a graph is an alternating series of vertices and edges beginning and ending with a vertex in which each edge is incident with the vertex immediately preceding it and the vertex immediately following it.A \textit{trail} is a walk in which all edges are distinct. A \textit{path} is a trail in which all vertices are distinct. A path has a \textit{Sequences of Vertices}. In a search tree, a \textit{Discrete Partition} or \textit{Individualization-Refinement Path} is a path which starts at the root and ends in a leaf \textbf{[1]}.

  A subset $B$ of a group  is called generating set, if the smallest subgroup containing the subset  is the group $S_n$ itself. We write, $\langle B \rangle =S_n$.  A generating set is called minimal generating, if the set  does not properly contain any generating set.

\section{Overview}

We define  graphs $G,H$ as \textit{regular, connected,  } and not \textit{locally triangle free}. If $G \simeq H$ then $H$ must have the same structure like $G$. We will rearrange $G$ according to \textbf{2.1}. This rearrangement will split $G$ in to vertex set $G_1, G_2.... G_x$ . In this paper, Graphs  are not  locally triangle-free. We will construct a search tree  $\mathcal{T}_k$  for $H_k$. The search tree  $\mathcal{T}_k$ will provide a set of permutations which will create the set $\beta_k$. We will construct generating set of automorphsim  from $\beta_k$. The algorithm uses the same technique to find automorphism  as [6].

\subsection  {\textit{Rearrangement} of a Graph:}

Consider any vertex of $G$, say $v$. We label $v$ with the integer $n$ (put $n$ on it, as label). We use subscript  $n$ to denote the $n$ labeled vertex and write $v_n$. Now, construct   neighbourhood induced sub-graph   $N_G[v_n]$ (closed neighbourhood).  We rewrite $ N_G[v_n]$ as $ G_1$ i.e. $ G_1=N_G[v_n]$. Now, select any unlabeled vertex from $G_1$  and label it with $(n-1)$. The vertex with label $(n-1)$ would be denoted as $v_{(n-1)}$. 
Now, construct    induced sub-graph $ N_{G_{1}}(v_{(n-1)})$ and label any vertex of it as $v_{(n-2)}$ .

We repeat  the above procedure again.  Label  an unlabeled vertex (i.e. an unlabeled vertex of $G \setminus G_1$) as $v_{(n-3)}$ and obtain $G_2$ based on adjacency (neighbourhood) of veretex $v_{(n-3)}$. Using the same procedure, we would be able to obtain $G_3, G_4, ... G_{(n/2)}$ subgraphs and to  label all vertices of $G$. Thus $G$ could be splitted into  $(n/3)$ subgraphs. 

There is a bijection from the vertex set of $G$ to a set $L= \{ 1,2..n\}$ of labels.

Now, let us define, a $n$-tuple (a sequence or ordered list of $n$ vertices) $w_G$ where vertices are ordered according to their labels (e.g. vertex labelled with label $1$ is in $1^{st}$ position in $w_G$).

$w_G$ is our  desired  arrangement (ordered) of graph $G$. For some $k$, $G_k$ has vertices from $i_k$ to $j_k$ of $w_G$ where
$(i_k+2=j_k)$. Here,  $i_k$ is the starting position of $k^{th}$ subgraph $G_k$ in $w_G$ and $j_k$ is the ending position of $k^{th}$ subgraph $G_k$ in $w_G$.  Subscript $k$ is used to distinguish  $i, j$ for the $k^{th}$ subgraph $G_k$ from other subgraphs.
For example, if $k=1$, then $G_1$ has vertices from $(n-2)$ to $n$ in $w_G$, here $i_1=(n-2)$ and $j_1=n$.

\subsection  {Construction of \textit{Search Tree} for  Generating Set :} 
 We have rearranged $G$ (one of the two given graphs $G$ and $H$). Now, we will construct permutations for $H$ with respect to $G$. Conversely, permutations can be generated  for $G$ with respect to $H$ using procedure described in this subsection.

We label each vertex of $H$  uniquely  with elements from the same set $L= \{ 1,2,...n\}$. This labeling procedure is random. It must make sure  that there is a bijection from the vertex set of $H$ to the label set $L$.

$w_H$  is a $n$-tuple (a sequence of $n$ vertices) where vertices are ordered according to their labels . The  definition of $w_H$ is similar to $w_G$, except it is defined for graph $H$.

In $w_G$, position starts from left to right, so the  subgraph $G_k$ starts at $i_k^{th}$ position and ends at $j_k^{th}$ where $n> j_k > i_k$. 

We define  the subgraph  $H_k $ of $ H$,  which has consecutive vertices from $i_k^{th}$ position to $j_k^{th}$ position in  $w_H$. Here,  $i_k$ and $j_k$ have the same value as they had in $G_k$. So, $H_k$ has $3$ vertices too.

 If $G_k \simeq H_k$ then $\exists \pi_k$ such that $H_k^{\pi_k}=G_k$. Permutation $\pi_k$ moves  vertices of $H$ to the interval between $i_k^{th}$ position and $j_k^{th}$ position, so that  $H_k^{\pi_k}=G_k$. So, we construct a search tree  $\mathcal{T}_k $ for constructing all possible permutations which could be $\pi_k$.  We will  follow the construction method of $G_k$ described in \textbf{2.1}, when we construct all possible  permutations that could be   $\pi_k$.

Let us define a rooted tree $\mathcal{T}_k$ (for $H_k$) , its nodes  are labelled vertices of $H$.
The children of root will be all possible candidates (vertices) for $j_k^{th}$ position of $H$ which could be $v_{j_k}$ where $v_{j_k}$ is the ${j^{th}_k}$ vertex of $w_G$. Since  $\pi_k$ moves (or fixes) a vertex of $H$ to the $j_k^{th}$ position of $H$  such that $H_k^{\pi_k}=G_k$, consider all possible vertices that could be the  vertex $v_{j_k}$ of $G_k \subset G$. All $n$ vertices of $H$ could be the $j_k^{th}$ vertex of $G$. So, at level $1$ , the children of root would be all $n$ vertices of $H$. It  means, each node of $1^{st}$ level, has a unique label in $\mathcal{T}_k $. Thus, we have  a bijection form  $H$ to  the nodes of $1^{st}$  level of  $\mathcal{T}_k $. Let, $t_{k_1}$ is a node of level $1$ of $\mathcal{T}_k $.

Each node  of  $t_{k_1} \in \mathcal{T}_k $ is  related to a vertex, say $u_l$ in $H$ ($1 \leq l \leq n $).  All vertices that are adjacent to $u_l$ make a subgraph, say $\mathbf{A_H}$. 

The children nodes  of $t_{k_1} \in \mathcal{T}_k $ will be all vertices of subgraph $\mathbf{A_H}$. Repeat previous  procedure $\forall t_{k_1} \in \mathcal{T}_k $. Thus we obtain all nodes of the $2^{nd}$ level of $\mathcal{T}_k $.

We repeat the procedure for all $H_k$  graphs  until we find  all possible leaf node  of $\mathcal{T}_k $. Thus, we construct the search tree $\mathcal{T}_k $. Note that, each level represents a position in $w_H$ of $H$, for example, in $\mathcal{T}_k $ of $H_k$, $1^{st}$ level represents the $j_k^{th}$ position. We construct $x$ such trees.

The height of the $\mathcal{T}_k $ is $2$ .
Note that an  individualization-refinement path or discrete partition  of $\mathcal{T}_k $ is a permutation of $H_k \subset H$ (concept of \textbf{ [1]}). All such paths, i.e. permutations  create the set $\beta_k$. There will be total $x$ number of $\beta_k$.

For example, if a \textit{path} is $5,8,9$ (where $5,8,9$ represent the labeled vertices of $H$)  then,      
$\pi_k=\left(\begin{matrix}   j_k & (j_k-1)  & (j_k-2) &=i_k  \\5 & 8 & 9  \end{matrix} \right)$.

It means, $\pi_k$ is a permutation  that moves $5^{th}$ vertex of $w_H$ to $j_k^{th}$ vertex,$8^{th}$ vertex of $w_H$ to $(j_k-1)^{th}$ vertex, and $9^{th}$ vertex of $w_H$ to $(j_k-2)^{th}$ vertex.

\section{Propositions}

If $G \simeq H$, then $\exists \pi_k \in \beta_k$ such that $H_k^{\pi_k}=G_k$. For each subgraph $H_k$, we have found a set of permutations $\beta_k$(from \textbf{2.2}).  We would be able to construct the direct product $P$ such that $H^{p}=G$ \textit{if and only if} $H \simeq G$. If we fail to construct such $P$, it implies that $H \not \simeq G$. 

\textbf{Proposition 3.1 :} $|\beta_k|  < n^{3}$. 

\textbf{Proposition 3.2 :} Given two graphs  $G, H$ with $n$ vertices  each, deciding whether they are isomorphic is \textit{polynomial time equivalent} to  determining   generating sets of automorphism group of  graphs  $G, H$ .

\textbf{Proof:} See  $[3]$.$\blacksquare$

 So,to decide graph isomorphism of $G, H$,  it is sufficient to construct   generating sets of automorphism group of  graphs  $G, H$.

\textbf{Proposition 3.3 :} Let $S_n$ be the finite group of order $n!$, There is a subset  of elements of $S_n$ of size at most
$\log_2(n!)$ which generates $S_n$.

\textbf{Proof:} The proof is similar to the Lemma 1 of $[4]$ on page 3.$\blacksquare$

From now on, $G, H$ are adjacency matrices of graphs $G, H$ respectively. $H_k, G_k$ are blocks or sub-matrices of matrix $H, G$ respectively. The adjacency matrix of graph $H_k \cup  H_e$ is $M_{(k,e)}$ where   $M_{(k,e)} =\left( \begin{array}{ccc} H_e & R_{k,e} \\ R_{k,e}^{T} & H_k\\ \end{array} \right) $,  where, $R_{k,e}$ is the non symmetric sub-matrix of adjacency matrix $H$.  Here, $R_{k,e}$ represents  edges  between $H_k,  H_e$.  Similarly,  $S_{k,e}$ represents  edges  between $G_k, G_e$.
$$H = \begin{bmatrix}
H_{(x)} & R_{(x, x-1)} & R_{(x,x-2)} & \dots & \dots & R_{(x,1)} \\
R_{(x,x-1)} & H_{(x-1)} & R_{(x-1,x-2)} & \dots & \dots & R_{(x-1,1)} \\ \vdots & \vdots & \vdots & \ddots & \ddots & \vdots \\
R_{(x,1)} & R_{(x-1,1)} & R_{(x-2,1)} & \dots & \dots &H_{1} \end{bmatrix}$$
\textbf{Proposition 3.4 :}  Generating set of   automorphism group   of graph  $H$ can be constructed  in polynomial  time if $H$ is not locally triangle-free as defined above  .

\textbf{Proof:} An algorithmic proof  is presented here.
 At $1^{st}$ iteration -

Step 1. Construct all possible  direct product  $(\pi_1 \times  \pi_2)$ where $\pi_1 \in \beta_1$ and $ \pi_2 \in \beta_2$. 

There are $| \beta_1 | \times | \beta_2|  <  n^{9}$ direct products (permutations).  All these permutations (direct products) form set $\gamma_1$. Each element of $\gamma_1$ is a permutation that acts on graph $H_1 \cup  H_2$.

Step 2. Construct/find -

$\alpha_1 =\{ \pi \in \gamma_1 |  (M_{(1,2)}^{\pi}= M_{(1,2)}) \land ( R_{1,2}^{\pi} = S_{1,2}) \land  (H_1^{\pi} = G_1) \land  (H_2^{\pi} = G_2) \}$ 

$\alpha_1$ is the set of automorphisms of matrix $M_{(1,2)}$.        $|\alpha_1| < n^{9}$.There are two possible cases-

  Case 1:  If $|\alpha_1| =1$, then  for each $\pi_1 \in \beta_1$,  there is only one permutation
$\pi_2 \in \beta_2$. So, there could be maximum $n^{2}$ permutations in $\gamma_1$ but only one permutation could be included in $\alpha_1$.
  
 Case 2: If $|\alpha_1| >1$,  we would be able to construct a generating set $\mathcal{S}_1$ of an automorphism group of $Aut(M_{(1,2)})$ Note, that if $\exists \pi_a \in Aut(H)$ such that it acts on   vertices of $H_1 \cup H_2$, then  $  \pi_a \in \langle \mathcal{S}_1 \rangle =Aut(M_{(1,2)})$. So, when we construct direct product of  $\mathcal{S}_1$ and another set,   $\pi_a$ can be found in the resulting generating set. This concept is similar to \textit{extending an automorphism} described in [6]. Also, see Theorem 7, on page 31 of [5].The theorem showed how to obtain the automorphism group of an arbitrary graph from the intersection of a specific  permutation group with a direct product of symmetric groups. 

Step 3. Now, we construct the generating set  $\mathcal{S}_1$  from $\alpha_1$. This construction of generating set can be done in polynomial time (see [5], page 40, theorem 9). From proposition 3.3, we find that $|\mathcal{S}_1| \leq log(n!)$ . 
 $\mathcal{S}_1$ is the  generating set  of automorphism of $M_{(1,2)}$ .

Step 4. We start $2^{nd}$ iteration, for $\beta_3, \mathcal{S}_1$ (instead of $\beta_2$), $ M_{(2,3)}$  where $M_{(2,3)} =\left( \begin{array}{ccc} H_3 & R_{2,3} \\ R_{2,3}^{T} & H_2 \\ \end{array} \right) $.  We find $\gamma_2, \alpha_2$ repeating  steps $1,2$ and construct $\mathcal{S}_2$ (repeating  step $3$)  which  is the  generating set  of automorphism of $M_{(2,3)}$, i.e.  graph $H_1 \cup H_2 \cup H_3$. Note that,                    $|\mathcal{S}_2|  \leq log(n!)$ .

Step  5. We keep repeating above four processes, until we find the set $\mathcal{S}_{(x-1)} $  which  is the  generating set  of automorphism of graph $H_1 \cup H_2 \cup H_3 \dots \cup H_x=H $. Note that, $|\mathcal{S}_{(x-1)}|\leq  log(n!)$, since $ \langle \mathcal{S}_{(x-1)} \rangle= Aut(H) \leq S_n$.  $\blacksquare$

 \section{ Conclusion}

  We repeat the process of construction of $\mathcal{S}_{(x-1)}$ for graph $G$ and obtain set $\mathcal{R}_{(x-1)}$. Once we generate  generating sets of $G, H$, we can decide isomorphism betwen them (3.2). The algorithm does not solve graph isomorphism problem in polynomial time  if graphs are \textit{locally triangle-free}. If the sub-matrix of edges  is not a zero matrix then  the problem  reduces down to \textit{Bipartite Graph Isomorphism Problem}. We can use the same approach there as above. This should lead to a practical solution which is the core idea of  practical graph isomorphism \textbf{[1]}.

\end{document}